\documentclass[12pt]{article}
\usepackage{graphics}

\newcommand{\tr}{{\rm tr}}

\begin{document}
\title{Fractal asymptotics}
\author{
C. P. Dettmann\thanks{Department of Mathematics, University of Bristol,
University Walk Bristol BS8 1TW, UK}}

\maketitle

\begin{abstract}
Recent advances in the periodic orbit theory of stochastically perturbed
systems have permitted a calculation of the escape rate of a noisy
chaotic map to order 64 in the noise strength.  Comparison with the usual
asymptotic expansions obtained from integrals and with a previous 
calculation of the electrostatic potential of exactly selfsimilar
fractal charge distributions, suggests a remarkably accurate form for the
late terms in the expansion, with parameters determined independently from
the fractal repeller and the critical point of the map.  Two methods
give a precise meaning to the asymptotic expansion, Borel summation and
Shafer approximants.  These can then be
compared to the escape rate as computed by alternative methods.
\end{abstract}

\begin{quote}
Keywords: Asymptotic expansions, Borel summation, Pad\'e approximation,
cycle expansions, escape rates, fractals, maps, repellers,
stochastic perturbations.\\
PACS: 02.30.Mv 05.40.Ca 05.45.Df
\end{quote} 

\section{Introduction}
Fractal sets and measures appear naturally as invariant sets (respectively
measures) of many nonlinear dynamical systems.  Periodic orbit
theory~\cite{Cv}
provides an effective approach to computing useful properties such as
averages, Lyapunov exponents and dimensions, particularly when the fractal
corresponds to a nonattracting set (``transient chaos'') so that direct
simulation methods are harder to implement.  In the case of Axiom A
dynamics, the convergence of periodic orbit (or ``cycle'') expansions
can be spectacular, see Tab.~1 below.

Recent work has extended the theory to include chaotic systems perturbed
by external noise, motivated by realism (all physical systems are coupled
to unknown degrees of freedom) and smoothness (delta functions are replaced
by smooth distributions).  While several approaches have been attempted,
including Feynman diagrams~\cite{CDMV98} (by analogy with
quantum perturbation theory) and
smooth conjugations~\cite{CDMV99} (by analogy with 
classical perturbation theory), the most
computationally effective method~\cite{CSPVD} has been to represent 
the stochastic
evolution (Fokker-Planck) operator by a matrix in a basis of polynomials
about each periodic orbit, truncated to finite size using the fact that
the elements decay exponentially, that is, the eigenfunctions are very
smooth.

The previous work~\cite{CSPVD} computed the escape rate of a stochastically
perturbed map
to order 8 in the noise strength.  In this paper the same matrix method,
extended to high (60 digit) precision arithmetic, is used to compute the
escape rate to order 64 in the noise strength.  With this number of
coefficients, it is meaningful to consider the asymptotic form of the
late terms, the subject of this paper.

We will discover that to unlock the secrets of the noise expansion will
require insights not only from the classical theory of asymptotic
expansions, but also from more recent analytic calculations involving
fractals.  The relatively small number of numerical coefficients is
compensated by their high precision, allowing a reliable fit to a functional
form involving several parameters.  Comparing two interpretations of
the series, Borel summation and Shafer (generalised Pad\'e) approximation
to the ``exact'' function, we will find that exponentially small corrections
will need to be considered by a future theory.

At this point we note a few other relevant works.  Contour integration
methods~\cite{PVVSD}
have obtained the asymptotic form of noise coefficients for fixed
periodic orbit length, however this does not directly determine the
noise expansion of the escape rate since the latter requires successively
longer orbits for higher noise corrections.
Direct integration~\cite{D99} has shown numerically that the
cumulant expansion on which the cycle expansions are based is valid for
strong as well as weak noise.  Stochastically perturbed dynamical systems
constitute a vast field, applying many methods other than periodic orbit
theory.

Section 2 outlines the previous theory and methods needed to understand
the results and their interpretation.  For space reasons, readers
interested in the full details are referred to the original works.
Section 3 gives the coefficients, the logic used to fit them to a 
particular functional form, and the Borel summation or Shafer approximation
needed to assign a precise meaning to the asymptotic expansion.  Final
discussion is given in section 4.

\section{Preliminaries}
\subsection{Asymptotic series}\label{s:asym}
This subsection gives the background for asymptotic series.  A very readable
review of this subject and its applications is given by Boyd~\cite{Boyd}.
Singular perturbations of integrals or differential equations,
such as perturbative approaches to physical problems, frequently have power
series expansions of the form
\begin{equation}\label{e:asym}
\sum_m(m+\alpha)!\left(\frac{x}{x_0}\right)^mM_m
\end{equation}
where $\alpha$ is often an integer or half integer, and $x_0$ (or
rather its inverse) is called the {\em singulant}, and is related to the
nearest critical point of an integrand. $M_m$ is called
the {\em modifying factor} and tends to a constant as $m\rightarrow\infty$;
it contains all the slower varying functional behaviour.

Such series diverge for all $x\neq 0$, and a number of methods have been
employed to make sense of them.  The simplest (albeit discontinuous)
is to truncate the sum at its smallest term.  An alternative, which Boyd
states as usually the most computationally efficient, is to replace the
series by its Pad\'e approximants, more specifically the Shafer
extension~\cite{Shafer}
in which the function is written as the solution of a quadratic equation
with polynomial coefficients; the coefficients of the polynomials are found by
a set of (typically ill-conditioned) linear equations.  

While Pad\'e methods give results on average as good as any other method
to this order, an alternative, Borel summation, offers the possibility
of systematic exponentially accurate (``hyperasymptotic'') corrections.
These methods (for example Berry and Howls~\cite{BH}) start from
the Borel summation method as formulated by
Dingle~\cite{Di}, which we follow. The latter retains the decreasing terms,
then performs Borel summation on the divergent ``tail''.  Borel summation is
an approach in which the factorial is replaced by its integral
representation~\cite{AS},
and the sum and integral are interchanged.  In the present context,
where the terms are all the same complex phase, this leads to a pole in
the path of integration (corresponding to a ``Stokes line'');
the constraint that
the result must be real then indicates that the principal value of the
integral should be taken.  Dingle also shows how to write the Borel summed
expression in terms of a few standard functions (related to incomplete
gamma functions~\cite{AS}) which he calls {\em terminants};
we use his method based on the forward difference expansion.
More details can be found in Dingle's book~\cite{Di}. 

For the bulk of this paper, we will need only Eq.~(\ref{e:asym}), in order
to fit the parameters, and generalise it slightly.  Once the form for the
coefficients has been established, the Shafer and Dingle methods will be
applied, and compared with the ``exact'' numerical result; the Shafer method
gives marginally more accurate results.  Since the coefficients are fitted,
not known exactly, and our calculation does not attempt to identify
exponentially small corrections (although this
should be possible in the future), we cannot apply the more detailed
hyperasymptotic methods of Berry and Howls~\cite{BH} and others (refer to
Ref.~\cite{Boyd} for extensive references).  Nevertheless,
we succeed in computing a function accurate to remarkably high values of
the noise strength (perturbation parameter).

\subsection{Exactly self-similar fractals}\label{s:cantor}
This subsection gives the background for analytic expansions pertaining
to exactly selfsimilar fractals, giving additional clues to the
value of the parameter $\alpha$ in Eq.~(\ref{e:asym})
and generalisations of this equation that we might expect for the noise
expansion in later sections.  The first paper to introduce asymptotic
methods applied to a fractal problem~\cite{BGM}
discussed Julia sets, but is more technically involved than the exactly
selfsimilar fractals in Refs.~\cite{DF1,DF2}.  The discussion in
this section is based on Ref.~\cite{DF1}.  

The middle-third Cantor set consists of two copies of itself scaled down
by a factor of three, thus it has dimension $d=\ln 2/\ln 3$ according to
many definitions.  The uniform measure on the set located between
$x=\pm1/2$ satisfies the relation
\begin{equation}
\int f(x)d\mu(x)=\frac{1}{2}\int\left[f\left(\frac{x-1}{3}\right)+
f\left(\frac{x+1}{3}\right)\right]d\mu(x)
\end{equation}
for arbitrary smooth function $f(x)$,
which together with the definitions of the electrostatic potential
\begin{equation}
V(x)=\int\frac{d\mu(x)}{|x-x'|}
\end{equation}
and the moments
\begin{equation}
C_n=\int x^n d\mu(x)
\end{equation}
lead after several steps~\cite{DF1} to the following expression for the
potential near the edge of the fractal,
\begin{equation}
V(1/2+\xi)=\xi^{d-1}\sum_{p=0}^{\infty}a_p\cos\left(\frac{2\pi p}{\ln 3}
\ln\xi+\phi_p\right)+\sum_{p=0}^{\infty}b_p\xi^p
\end{equation}
where $a_p$, $b_p$ and $\phi_p$ are known series given in terms of the $C_n$,
which are in term given as explicit rational numbers with a well understood
asymptotic form.
The $a_p$  have a large exponential decay rate for example $a_0=1.7685$,
$a_1=7.04977\times 10^{-8}$, $a_2=6.7575\times 10^{-17}$.  The 
oscillatory terms come from the $p\neq 0$ solutions
\begin{equation}
d=\frac{\ln 2}{\ln 3}+\frac{2\pi i p}{\ln 3}
\end{equation}
of the equation $3^d=2$ defining the dimension.  Such ``complex dimensions''
of fractals appear in a number of physical applications~\cite{SS}.

Even though the context is different from that of the noise corrections
considered below in Sec.~\ref{s:osc}, in particular the expansion for
the potential has a finite radius of convergence while that of the noise
is divergent, we may conjecture that there are a number of general
principles applying to fractal expansions.  In particular, the leading
order of the expansion is not a single term but a sum of terms containing
a variable (here $\xi$) raised to powers $\alpha+ip\beta$ for all integers $p$,
where $\alpha$ and $\beta$ are determined by properties of the fractal:
\begin{enumerate}
\item The parameter $\alpha$ (here $d-1$) is related to the dimension of the
fractal.
\item The parameter $\beta$ (here $2\pi /\ln 3$) is determined by the spatial
scaling factor $3$ of the fractal.
\item Due to the rapid decay of the oscillatory coefficients, accurate
results may be obtained by considering only one or two of these terms,
ie $|p|\leq 1$.
\end{enumerate}
Note that in the absence of the imaginary $\beta$ terms (perhaps a ``nonfractal
limit''), the $\alpha$ in this section corresponds to the $\alpha$ in
Eq.~(\ref{e:asym}) since the latter effectively has $\alpha$ in the exponent:
$(m+\alpha)!\sim m! m^\alpha$ as $m\rightarrow\infty$.

\subsection{Periodic orbit theory of stochastically perturbed maps}
\label{s:noise}
This subsection gives a background to the numerical coefficients
discussed in the main part of the paper.  The periodic orbit theory~\cite{Cv}
allows the computation of long time properties such as averages,
Lyapunov exponents, and correlation functions from periodic orbits.
The required properties are related to the leading eigenvalue(s) of
an evolution operator (defined below), which, in the most rapidly
convergent formulation, are computed using determinants.  The determinant
is expressed as an expansion in traces, and the traces are expressed in
terms of periodic orbits.

Consider the discrete time dynamics described by
\begin{equation}\label{e:dyn}
x_{n+1}=f(x_n)+\sigma\xi_n
\end{equation}
where $\sigma$ is the noise strength and $\xi_n$ are instances of an
uncorrelated random variable.  In the present case $x$ is a real number,
but higher dimensional generalisations are straightforward, at least in
the deterministic case $\sigma=0$.  Continuous time dynamics can also
be considered; the stochastic version has been discussed in Ref.~\cite{G02}.
The probability density $\rho(x)$ evolves according to
\begin{equation}
\rho_{n+1}(y)=({\cal L}\circ\rho_n)(y)=
\int\rho_n(x)\delta_\sigma(y-f(x))dx
\end{equation}
where $\cal L$ as defined above is the (discrete time Fokker-Planck)
evolution operator.  The
noise distribution $\delta_\sigma(z)$ is an arbitrary function
of standard deviation $\sigma$; it reduces to a Dirac delta in the
deterministic case $\sigma=0$.  We compute the trace
\begin{equation}
\tr{\cal L}^n=\int dx_0 dx_1\ldots dx_{n-1} \delta_\sigma(x_1-f(x_0))
\delta_\sigma(x_2-f(x_1))\ldots\delta_\sigma(x_0-f(x_{n-1}))
\end{equation}
which is an $n$-dimensional integral.  In the deterministic case, it
reduces to a sum over periodic points $x$ satisfying $f^n(x)=x$ of the
relevant Jacobian.  The weak noise theory is effectively a stationary
phase approximation in which the leading order behaviour is given by
the deterministic limit, with corrections determined by higher derivatives
of the map $f(x)$ evaluated at the periodic points.  Exponentially small
corrections are obtained by considering local extrema, which are
given by the ``generalised periodic orbits'' of~\cite{PVVSD}, that is,
periodic orbits of the extended system $(x,p)\to(f(x)+p/f'(x),p/f'(x))$;
these will be discussed later.

The characteristic determinant is
\begin{equation}\label{e:cum}
0=\det(1-z{\cal L})=\exp\tr\ln(1-z{\cal L})=1-z\tr{\cal L}
-\frac{z^2}{2}(\tr{\cal L}^2-(\tr{\cal L})^2)+\ldots
\end{equation}
where we define the determinant in terms of its expansion in powers of the
inverse eigenvalue $z$; this is consistent with our desire for the largest
eigenvalue (smallest $z$).  The eigenvalue itself can be obtained by
truncating the above equation at some $z^n$, which requires computing
periodic orbits up to length $n$, and solving numerically for the first
zero.  The leading eigenvalue $\nu=z^{-1}$ has a direct interpretation:
the quantity $\gamma=-\ln\nu$ is the escape rate, that is, a uniform
distribution of initial conditions leads to a number proportional to
$\exp(-\gamma n)$ remaining at long times $n\gg 1$.  Other useful
quantities can be obtained by small modifications of the method, for example
weighting the evolution operator using the function for which an average
is desired.

We note that because the eigenvalue is not directly expressed as an integral,
rather an expansion of integrals or the infinite dimensional limit of an
integral, the Dingle theory discussed in Sec.~\ref{s:asym} does not
strictly apply.  In principle (although not in practice) an infinite number
of critical points are required to determine the eigenvalue.

There have been two analytical approaches to evaluating the integral
for the trace in the stochastic case, in particular Feynman
diagrams~\cite{CDMV98} and smooth conjugations~\cite{CDMV99}.  These give
the trace explicitly in terms of the derivatives of $f(x)$ at the periodic
points, but have been applied only up to order $\sigma^4$.  A more numerical
approach was used in Ref.~\cite{CSPVD} to obtain coefficients up to $\sigma^8$,
and it is the latter approach which is used in this paper.

The evolution operator is expressed in an explicit polynomial basis
(in contrast to the usual situation in periodic orbit theory, where all
calculations are kept independent of the basis), defined in the
vicinity of each periodic point.  Truncating the representation to a
finite matrix is justified since the eigenfunction is very smooth, leading
to exponential decay of the matrix elements.  All quantities are expanded
in powers of the small parameter $\sigma$, and the full trace is obtained
as a sum over periodic points, to each order in $\sigma$.  Finally the
leading eigenvalue is obtained as a formal power series
\begin{equation}\label{e:num}
\nu(\sigma)=\sum_{m=0}^\infty \nu_{2m}\sigma^{2m}
\end{equation}
where odd powers vanish by the symmetry of the Gaussian noise distribution
used.  Further details of the calculation are given in Ref.~\cite{CSPVD}.

\section{Results}
\subsection{Numerical details}
In the previous paper~\cite{CSPVD} the results were limited to
$\sigma^8$ since high
precision is required (the cumulant expansion~(\ref{e:cum}) involves
many cancellations), and commercial high precision mathematical packages
required too much memory and time.  The results of this paper were
achieved by code written in C, involving 60 digits
precision, a maximum matrix size of 200 (the largest matrices are required
for the shortest orbits) and periodic points up to $n=10$.  The matrix size
for each orbit length and the maximum length were determined adaptively;
the precision was estimated conservatively using the results of shorter
calculations.  Here, as in previous calculations~\cite{CDMV98,CDMV99,CSPVD},
the noise is Gaussian, and the map appearing in~(\ref{e:dyn}) is
\begin{equation}\label{e:map}
f(x)=20\left[\frac{1}{16}-\left(\frac{1}{2}-x\right)^4\right]
\end{equation}

\begin{figure}
\begin{picture}(290,240)
\put(100,20){\scalebox{0.7}{\includegraphics{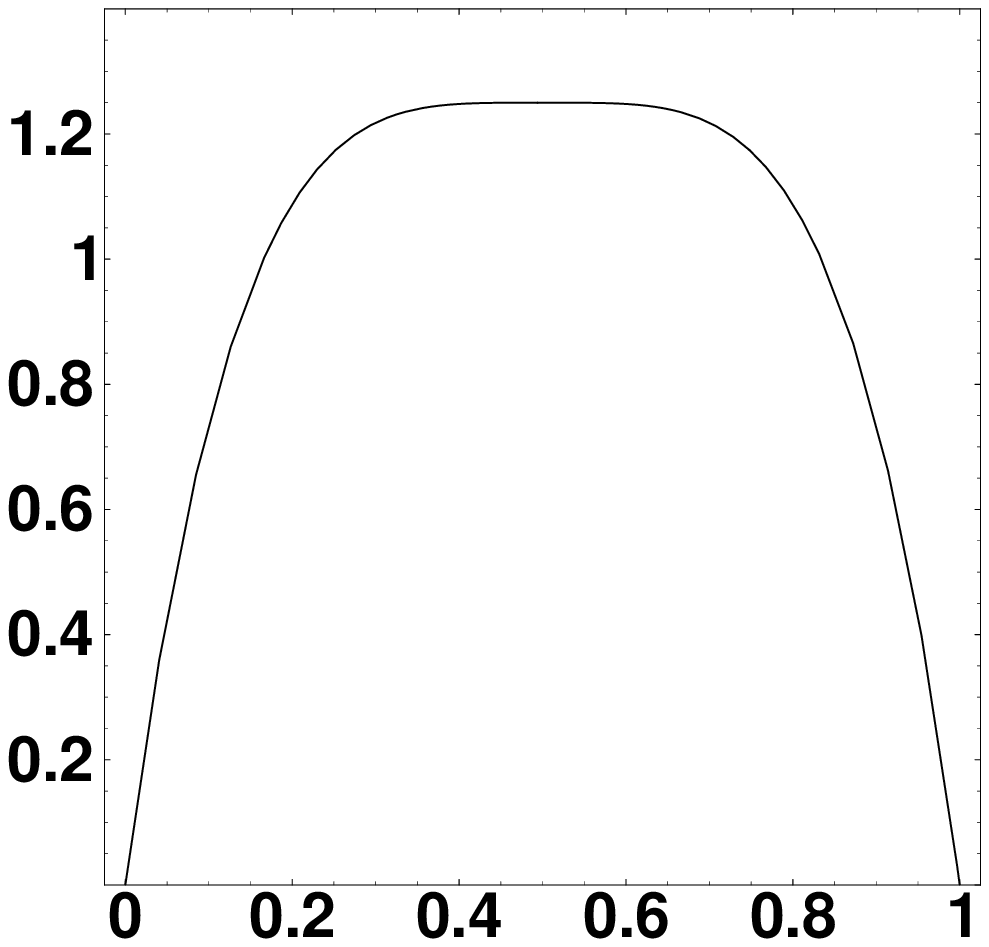}}}
\put(200,0){\Large $x$}
\put(60,120){\Large $f(x)$}
\end{picture}
\caption{The map~(\protect\ref{e:map}) appearing in~(\protect\ref{e:dyn}).
}
\end{figure}

This map is Axiom A with complete binary symbolic dynamics, so the rate
of convergence of the cycle expansion with orbit length is super-exponential,
both for $\nu_0$ (the deterministic case) and the noise corrections.
Examples of this convergence (noted in the previous studies) are
given in Tab.~\ref{t:conv}.  The results for orbit length $n=10$, expressed
as the logarithms of the (always positive) $\nu_{2m}$ are given in
Tab.~\ref{t:res}.

\begin{table}
\begin{tabular}{|r|r|r|r|r|r|}
\hline $n$&$\nu_0$&$\nu_2$&$\nu_4$&$\nu_{64}$\\\hline
1&0.307735902965&0.421227543767&2.15906608736&$1.397115735\times 10^{53}$\\
2&0.371401067274&1.421640613096&32.97365355137&$5.001186917\times 10^{75}$\\
3&0.371109569907&1.435552381965&36.32563272348&$2.001067045\times 10^{80}$\\
4&0.371110995255&1.435811262322&36.35837768356&$2.651047356\times 10^{80}$\\
5&0.371110995235&1.435811248197&36.35837123374&$2.660918038\times 10^{80}$\\
6&0.371110995235&1.435811248197&36.35837123384&$2.660918375\times 10^{80}$
\\\hline
\end{tabular}
\caption{\label{t:conv}
The noise coefficients of the eigenvalue, as defined in
Eq.~(\protect\ref{e:num}), calculated using periodic orbits up to length $n$.}
\end{table}

\begin{table}
\begin{tabular}{|r|r|r|r|r|r|}
\hline $2m$&$\ln\nu_{2m}$&$2m$&$\ln\nu_{2m}$&$2m$&$\ln\nu_{2m}$\\\hline
0&-0.99125408258905&22&49.59463191979195&44&117.84381872333873\\
2&0.36173001922727&24&55.44039700516978&46&124.39362242853359\\
4&3.59342447275687&26&61.37352232130161&48&130.98895554172700\\
6&7.63842801043723&28&67.38731548909859&50&137.62787534158868\\
8&12.15107844830820&30&73.47605210071374&52&144.30858981612283\\
10&16.97610609587220&32&79.63476897706259&54&151.02944299196041\\
12&22.03308958456985&34&85.85911504876676&56&157.78889107050155\\
14&27.27470590323921&36&92.14524039569617&58&164.58550964455421\\
16&32.67034394383728&38&98.48971155485100&60&171.41796376702617\\
18&38.19872657460701&40&104.88944552922095&62&178.28500846576056\\
20&43.84418495656653&42&111.34165748821478&64&185.18547875658766\\\hline
\end{tabular}
\caption{\label{t:res}
The natural logarithm of the noise coefficients of the eigenvalue,
as defined in Eq.~(\protect\ref{e:num}).  All $\nu_{2m}$ are positive.} 
\end{table}

\subsection{Fitting $\nu_{2m}$ to the form~(\protect\ref{e:asym})}\label{s:fit}
We immediately note that, while the coefficients $\nu_{2m}$ converge
very rapidly with orbit length $n$, they diverge with order $m$.  This
is not surprising since the noise expansion of the eigenvalue is effectively
a stationary phase expansion of an integral, albeit with an infinite number of
critical points.  See Secs.~\ref{s:asym},~\ref{s:noise}.  A little curve
fitting leads to the very approximate form
\begin{equation}\label{e:approx}
\nu(\sigma)=\sum\nu_{2m}\sigma^{2m}\approx
\sum m!32^m\sigma^{2m}
\end{equation}
that is, $\alpha=0$ and $\sigma_0=32^{-1/2}$ by comparison
with~(\ref{e:asym}).  We would like to know the form of the coefficients
more precisely than this, and in particular predict them from other
information about the dynamics.

From the general theory of asymptotic expansions (Sec.~\ref{s:asym})
the singulant $\sigma_0$ is somehow related to the distance between
the critical point we are expanding around (periodic orbits on the
fractal repeller) and the nearest critical point.  In the present
situation, if we consider the critical point of the original
map~(\ref{e:map}) at $x_c=1/2$, and ask for the probability for returning
to the repeller, of which the most accessible point is $x_r=1$,
to first approximation this is
\begin{equation}
\exp\left[-(x_r-f(x_c))^2/2\sigma^2\right]
=\exp\left[-1/32\sigma^2\right]
\end{equation}
which indeed gives the 32.  The reason that the singulant $\sigma_0$
should be related to the coefficient of an exponential is that this
exponentially small quantity (for small $\sigma$), is, up to slower
varying factors, the magnitude of the smallest term in the expansion,
and hence the order at which exponentially small hyperasymptotic
terms might contribute.

The above expression assumes that the transition from the critical point
$x=1/2$ to the repeller $x=1$ takes place in a single step.  Actually,
for sufficiently small noise, longer trajectories may be more likely.
The probability is of the form $\exp[-\sum(x_{n+1}-f(x_n))^2/2\sigma^2]$
which can be maximised over all trajectories starting at the critical point
$x=1/2$ and reaching the repeller in the infinite time limit.
The result is\\ $\sigma_0=32.31850341240166^{-1/2}$ for the
trajectory $\{0.5, 1.00244613635157, -0.00024587488150,$
$-2.460023246\times 10^{-5},$ $-2.46015082\times 10^{-6},$
$-2.4601636\times 10^{-7}, -2.460165\times 10^{-8},\ldots,0\}$.  The approach
to zero is geometric with ratio $f'(0)^{-1}=1/10$. 

Another interpretation of this orbit is as the infinite length limit
of a sequence of generalised periodic orbits (Sec.~\ref{s:noise}),
responsible for exponentially small corrections to the traces; see
later discussion in Sec.~\ref{s:Borel}.  For the rest of this paper,
we assume that the 32 in Eq.~(\ref{e:approx}) is replaced by the adjusted
value $\sigma_0^{-2}=32.3185\ldots$. 

\begin{figure}
\begin{picture}(290,250)
\put(30,0){\scalebox{1.4}{\includegraphics{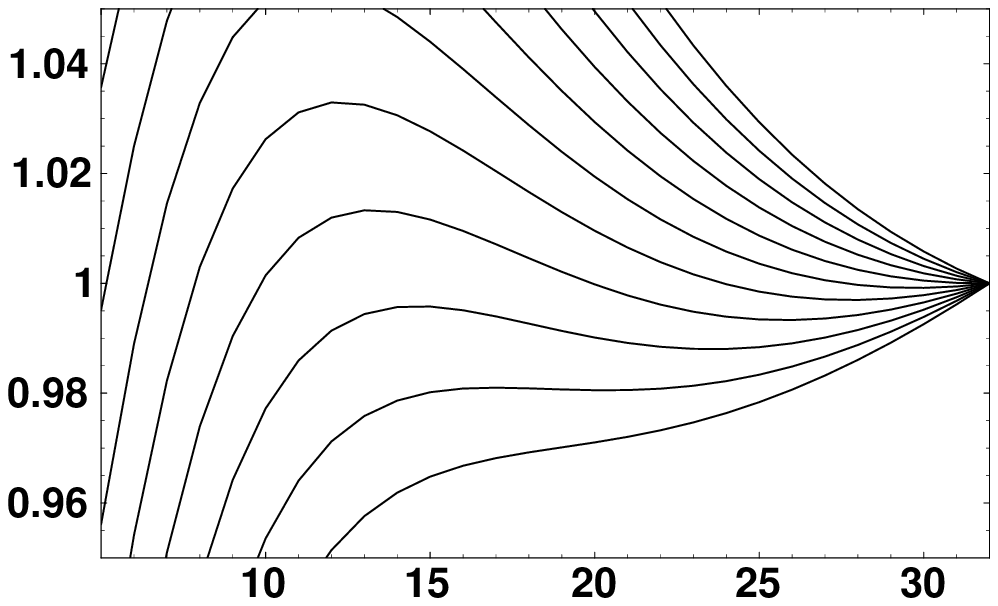}}}
\put(240,0){\Large $m$}
\put(0,135){\LARGE $\frac{M_m}{M_{32}}$}
\put(360,195){\large $\alpha=-1.4$}
\put(360,85){\large $\alpha=-1.2$}
\end{picture}
\caption{Attempt to fit the $\nu_{2m}$ to a single asymptotic series of
the canonical form.  See Eq.~(\protect\ref{e:rough}).
For none of the $\alpha$ do the coefficients asymptote to
a convincing horizontal line.\label{f:rough}}
\end{figure}

The only remaining parameter to be fitted then seems to be the $\alpha$
appearing in~(\ref{e:asym}).  With this in mind, we plot the quantity
\begin{equation}\label{e:rough}
M_m=\frac{\nu_{2m}\sigma_0^{2m}}{(m+\alpha)!}
\end{equation}
normalised to the highest order $M_{32}$ for various $\alpha$
in Fig.~\ref{f:rough}.  

As Fig.~\ref{f:rough} shows, the $M_m$ do not approach
a constant for any value of $\alpha$.  Even for the $\alpha\approx-1.3$
at which the curve is roughly horizontal for the largest $m$, the curvature
(as measured by the second derivative) is still large.  There appears to
be some oscillatory behaviour evident.

\subsection{Oscillations from complex exponents}\label{s:osc}
At this point we recall the discussion of Sec.~\ref{s:cantor}, in particular
the three observations relating to fractal expansions at the end of that
section.  If we can identify $\alpha$ in some way with the dimension of the
fractal (at least in the Cantor set example), the oscillations noted at
the end of the previous section appear naturally from a sum (over $p$) of
terms of the form $(m+\alpha+ip\beta)!$ for integer $p$.  Since the real
part of $\alpha+ip\beta$ is a constant $\alpha$, all of these terms are
of the same order in the large $m$ limit.

We return later to first observation in Sec.~\ref{s:cantor}, that is,
the question of whether $\alpha$ is related to one of the fractal
repeller's many dimensions, and leave it as a free parameter for
the present. The second observation suggests that we look at the
spatial scaling factor of the fractal repeller.  The orbit discussed in
the previous section reaches the fractal repeller at $x=0$; at this point
the repeller is selfsimilar with a scaling factor of $f'(0)=10$; the
same scaling factor that appears in the critical orbit of the previous
section.  The second observation of Sec.~\ref{s:cantor} then suggests
$\beta=2\pi/\ln10$, which matches the oscillations well (see below).
The third observation suggests that only small $p$ may lead to sizable
contributions, hence we will ignore $|p|\geq 2$ and only include the
real $\alpha$ and a single pair of complex conjugates $\alpha\pm i\beta$.

\begin{figure}
\begin{picture}(290,250)
\put(30,0){\scalebox{1.4}{\includegraphics{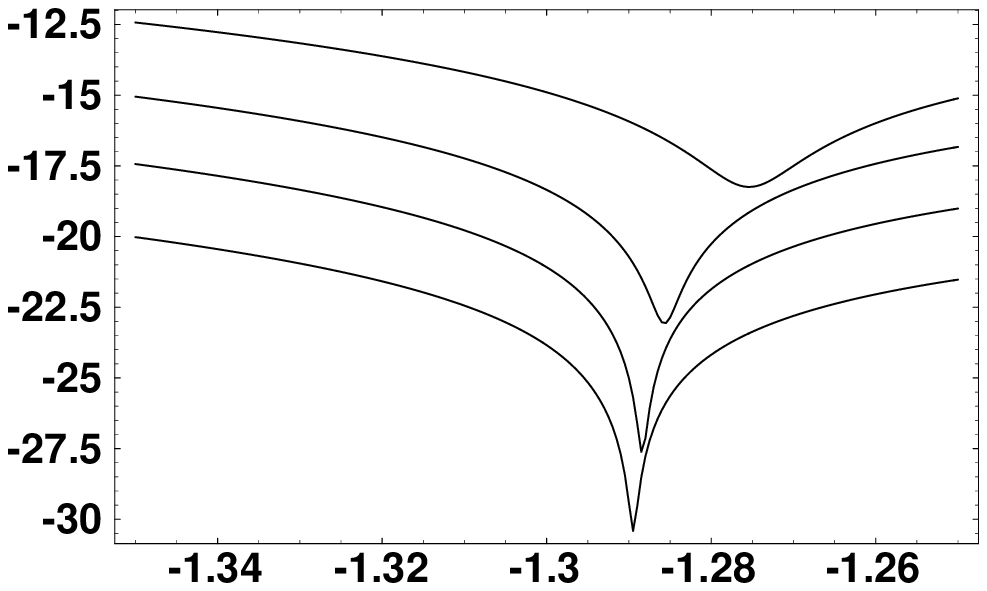}}}
\put(240,0){\Large $\alpha$}
\put(0,135){\Large $\ln\chi^2$}
\put(360,220){\large $m_{\rm min}=8$}
\put(360,110){\large $m_{\rm min}=20$}
\end{picture}
\caption{Effectiveness of the fit to the $M_m$ in
Eq.~(\protect\ref{e:rough}) using the function in Eq.~(\protect\ref{e:fit}).
Note the dramatic spike at $\alpha=-1.290$.
\label{f:chi}}
\end{figure}

\begin{figure}
\begin{picture}(290,250)
\put(30,0){\scalebox{1.4}{\includegraphics{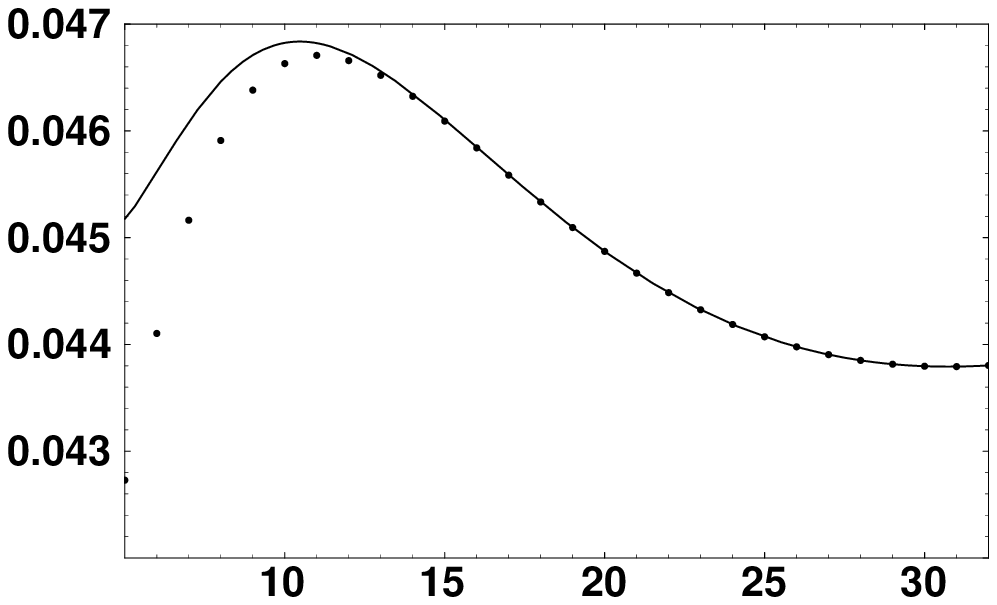}}}
\put(240,0){\Large $m$}
\put(-20,135){\Large $M_m$}
\end{picture}
\caption{Optimal fit to the $M_m$ in Eq.~(\protect\ref{e:rough}) (dots)
using the function in Eq.~(\protect\ref{e:fit}) (curve), with $\alpha=-1.290$,
$c0=0.045514$ and $c_1=0.000958+0.000185i$.  For most of the range of $m$
the difference between the two is much smaller than the scale visible
on this plot.
\label{f:fit}}
\end{figure}

We thus fit the data, as represented by the $M_m$ of Eq.~(\ref{e:rough})
to the function
\begin{equation}\label{e:fit}
c_0+c_1\frac{(m+\alpha+2\pi i/\ln 10)!}{(m+\alpha)!}
+c_1^*\frac{(m+\alpha-2\pi i/\ln 10)!}{(m+\alpha)!}
\end{equation}
where the $*$ indicates complex conjugation, $c_0$ is a real fit parameter
and $c_1$ is a complex fit parameter.  Note that dividing through by
$(m+\alpha)!$ in Eq.~(\ref{e:rough}) permits a linear (hence more reliable)
fit for $c_0$ and $c_1$. 
This fit is made for a
range of values of $\alpha$, and for $m_{\rm min}<m<32$ with various
$m_{\rm min}$.  The error (in the least squares sense) $\chi^2$ is
given in Fig.~\ref{f:chi}, which shows an improvement limited by the
precision of the results (14 decimal places in Tab.~\ref{t:res}).
From the optimal fit, we have $\alpha=-1.290$, corresponding to
$c_0=0.045514$ and $c_1=0.000958+0.000185i$.  This fit is shown in
Fig.~\ref{f:fit}.

We now return to the question of whether $\alpha$ is related to a dimension
of the fractal repeller.  The Renyi generalised dimensions $D_q$ of the
repeller are straightforward to compute using the usual periodic orbit
theory of deterministic systems~\cite{Cv}.  The results are given in
Tab.~\ref{t:dim}, but do not seem to exactly match our fitted value
$\alpha=-1.290$.  The closest is $D_\infty$ (or rather $-1-D_\infty$),
however this corresponds to
the most stable periodic orbit which is the fixed point at $x\approx0.871$,
not the (most unstable) point $x=0$ used in the above calculation of the
complex exponents.

\begin{table}
\begin{tabular}{|c|c|c|c|c|c|}\hline
$q$&$-\infty$&0&1&2&$\infty$\\\hline
$D_q$&0.5695&0.4007&0.3872&0.3757&0.2957\\\hline
\end{tabular}
\caption{Renyi dimensions of the fractal repeller, computed using
periodic orbit theory.\label{t:dim}}
\end{table} 

\subsection{Borel summation}\label{s:Borel}
To conclude the analysis of the results, we apply the Shafer (generalised
Pad\'e) and Dingle (Borel summation)
methods discussed briefly in Sec.~\ref{s:asym} and compare the results with
the exact eigenvalue $\nu(\sigma)$ computed using the discretized
eigenfunction method of Ref.~\cite{CDMV98} which is accurate to about
six digits.  Note that the full power of the Borel summation applied
to the full computed series up to order $\sigma^{64}$ is not (currently)
testable, since this is the smallest term when it (and hence the
expected errors) is of order $(2e)^{-64/2}\approx 10^{-24}$.
Instead, we will consider
quite large values of $\sigma$, where the smallest term is very close
to the beginning of the series.

The only slight extension of the Dingle approach discussed in
Sec.~\ref{s:asym} concerns the modifying factor $M_m$.  Since the asymptotic
series now has three components, with $p=0, \pm 1$, it is not clear whether
each component should have a modifying factor.  The point of view taken
here is that a single modifying factor $M_m$ is used for all three components.
This choice is pragmatic; while it is probably more natural to modify each
series separately, there is no method of extracting this information
from the numerical data. 

\begin{figure}
\begin{picture}(290,250)
\put(30,0){\scalebox{1.4}{\includegraphics{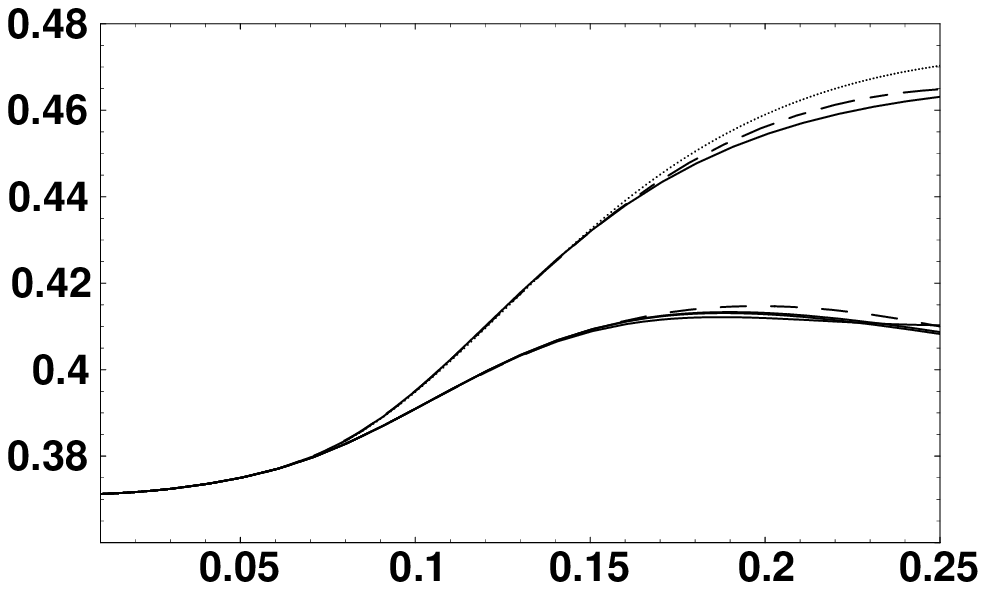}}}
\put(240,0){\Large $\sigma$}
\put(0,130){\Large $\nu(\sigma)$}
\end{picture}
\caption{Borel summation of the series (lower solid curves) and Shafer
approximant (dashed curve) together with an independent numerical
calculation of the eigenvalue (dotted line).
When an exponentially small function $0.09\exp(-\sigma_0^2/\sigma^2)$
is added to one of the former, the result is the upper solid line, and
the dot/dash line respectively.  These are much closer fits for larger
values of the noise.  Note that at the largest values of $\sigma$ shown here,
the asymptotic series is very rapidly divergent, with the minimum term given by
$1.44\sigma^2\approx0.09$.
\label{f:resum}}
\end{figure}

The results of the Borel summation are shown in Fig.~\ref{f:resum}.  The bunch
of lower solid curves are the Borel summed series, using five forward
differences (see Sec.~\ref{s:asym}) and from two to four terms of the
series before truncation.  Note that the Borel summed function is consistent
for relatively large $\sigma$, despite the obvious approximations
made in approximating the series by so few terms.  The dashed line is the
result of a Shafer (quadratic Pad\'e) approximation to the first
13 nonzero coefficients, as discussed in Sec.~\ref{s:asym}; the exact number of
coefficients fitted makes little difference.  These two interpretations of
the power series are quite consistent.  The dotted line is the true
eigenvalue, computed as in Ref.~\cite{CDMV98}.  It is extremely close
to the Borel summed series for $\sigma<0.08$, after which it is significantly
higher. 

The difference may be modelled by a function which is exponentially small
for $\sigma\rightarrow0$.  The most obvious candidate (but one of many) is
$C\exp(-\sigma_0^2/\sigma^2)$, which is roughly the magnitude of the
smallest term, and is also the first expected ``hyperasymptotic'' correction.
The remaining solid curve in Fig.~\ref{f:resum} gives one of the Borel summed
curves, plus this function with $C=0.09$.  The dot-dashed curve gives the
equivalent result for the Shafer approximant.  The result is a fit valid for
roughly $\sigma<0.16$, indicating that the exponential part of the
correction has the right form.

It is reasonable to identify this exponentially small correction as the
contribution of the generalised periodic orbit in Sec.~\ref{s:fit}.
However, there is actually a fractal set of such generalised periodic
orbits, starting from the critical point of the map and limiting to
each orbit on the repeller.  The orbit given in Sec.~\ref{s:fit} is
the most probable case, but other orbits have exponents which are
arbitrarily close.  The nontrivial task of summing these
contributions is left to a future paper. 

Incidentally, there exist rigorous results pertaining to Borel summation
of asymptotic series~\cite{Hardy}.  The observation of hyperasymptotic
corrections implies the presence of singularities in the complex noise
domain, however it is difficult to understand what physical consequences
this might have.

\section{Conclusion}
Periodic orbit theory of stochastic systems as presented in Ref.~\cite{CSPVD}
has been used to compute the escape rate of a stochastically perturbed map
to order 64 in the noise strength with sufficient precision to permit the
theory of asymptotic expansions to be applied.  Similar to a previous
calculation of an exactly selfsimilar fractal, complex exponents appear,
and the parameters in the expansion were found by a combination of
analytic arguments and curve fitting.  Although connections between
this and the previous calculation were at a phenomenological level, the
precise fit gives strong numerical evidence that the form of the expansion
is correct to this order.  Finally the Shafer approximation and Borel summation
were performed on the series and compared with the known escape rate function,
giving evidence for hyperasymptotic corrections of the
same order as predicted by the theory of asymptotic expansions and
independently by nonleading stationary points of the action.

In the future, the analytic connections proposed between the singulant
and the probability of returning to the repeller from the critical point,
and between the imaginary exponent and the spatial scaling factor of the
repeller, should be verified by calculations on a variety of different
systems.  The remaining tentative connection, between the real exponent
$\alpha$ and the $D_\infty$ dimension of the repeller would then either be
verified or contradicted.  The latter possibility is the most intriguing,
since in that case $\alpha$ could define a new ``noise dimension''. 

\section*{Acknowledgments}
The author is grateful for helpful discussions with Michael Berry,
Predrag Cvitanovi\'c, Giovanni Gallavotti, Jon Keating, Jonathan Robbins
and Niels S\o{}ndergaard.  This research was supported by the Nuffield
Foundation, grant NAL/00353/G.

\end{document}